# Social Internet of Things and New Generation Computing- A Survey

Hamed Vahdat-Nejad, Zahra Mazhar Farimani, Arezoo Tavakolifar

PerLab, Department of Computer Engineering, University of Birjand, Iran

{vahdatnejad, mazhar.zahra, tavakolifar.arezoo}_@birjand.ac.ir

## Abstract

Social Internet of Things (SIoT) tries to overcome the challenges of Internet of Things (IoT) such as scalability, trust and discovery of resources, by inspiration from social computing. This survey aims to investigate the research done on SIoT from two perspectives including application domain and the integration to the new computing models. For this, a two-dimensional framework is proposed and the projects are investigated, accordingly. The first dimension considers and classifies available research from the application domain perspective and the second dimension performs the same from the integration to new computing models standpoint. The aim is to technically describe SIoT, to classify related research, to foster the dissemination of state-of-the-art, and to discuss open research directions in this field.

**Keywords**- Social Internet of Things, Application domain, Computing model, Survey

## 1. INTRODUCTION

Nowadays, we live in a digital world where the number and variety of smart devices are increasing rapidly (Atzori, Iera, & Morabito, The Internet of Things: A survey, 2010). A smart object can be a sensor or any physical device that has the ability of sensing the environment, collecting data, connecting to the network and processing (Whitmore, Agarwal, & Da Xu, 2015). Some examples of these devices are smart phones, smart watches, smart TVs, medical and health devices, and vehicles (Stankovic, 2014). The integration of these devices through Internet connectivity introduces a new field called Internet of Things (IoT) (Al-Fuqaha, Guizani, Mohammadi, Aledhari, & Ayyash, 2015). In fact, the Internet of Things is composed of a global network of smart things that each of them has its own address and communicates with others based on the standard agreements ( Aggarwal, Ashish, & Sheth, 2013).

The Internet of Things offers smart services to control the things and user's environment by creating the connection between smart things (Jara, Bocchi, & Genoud, 2014). Nevertheless, the fulfillment of the Internet of Things has faced some challenges such as scalability, trust and discovery of resources, information and services (Miorandi, Sicari, De Pellegrini, & Chlamtac, 2013). Hence, Social Internet of Things (SIoT) has been recently proposed, with an idea that has originated from the social network (Atzori, Iera, & Morabito, SIoT: Giving a Social Structure to the Internet of Things, 2011; Atzori, Iera, Morabito, & Nitti, The Social Internet of Things (SIoT) – When social networks meet the Internet of Things: Concept, architecture and network characterization, 2012). The novel paradigm of SIoT has emerged to describe the world where everything around the humans can be cleverly comprehended and interacted (Zargari Asl, Iera, Atzori, & Morabito, 2013). Indeed, SIoT attempts to solve the problems of scalability, trust and discovery of resources, information and services by modeling the IoT as a social network. Hence, it provides a platform for better interactions between people and things (Girau, Martis, & Atzori, 2015; Nitti, Girau, Atzori, Iera, & Morabito, A subjective model for trustworthiness evaluation in the social Internet of Things, 2012; Nitti, Atzori, & Cvijikj, Network navigability in the social Internet of Things, 2014; Rabadiya, Makwana, & Jardosh, 2017; Ghane'i-Ostad, Vahdat-Nejad, & Abdolrazzagh-Nezhad, 2018). Furthermore, privacy protection technologies utilized in social networks can be employed to improve the IoT security (Xiao, Sidhu, & Christianson, 2015; Lin & Dong , 2018; Truong, Lee, Askwith, & Lee, 2017; Nitti, Girau, & Atzori, Trustworthiness Management in the Social Internet of Things, 2014). In the past years, SIoT has attracted the attention of many researchers.

Previously, several survey studies have examined the Internet of Things from various aspects such as architecture (Lee & Kwon, 2015). Similarly, the field of SIoT has been investigated in terms of the operating system in 2018 (Afzal, Umair, Shah, & Ahmed, 2018). However, the need for a research that investigates the existing SIoT systems in terms of the application domain and the way of integrating in the new computing models is felt. Available research studies on SIoT have been conducted in different domains and also they have exploited specific computing paradigms such as cloud or edge. This paper proposes a framework with two dimensions of application domain and computing paradigm for investigating the SIoT systems, and then considers the related research with respect to these dimensions. Finally, the conclusion remarks and future research directions are discussed.

After this introduction, the proposed framework and the specifications of the considered projects are investigated in section 2. In section 3, related projects are explored in terms of the application domain. Section 4 explains the computing paradigm used in the related projects, and finally the conclusion remarks and future research directions are discussed in section 5.

## 2. THE PROPOSED FRAMEWORK

In this research, the SIoT papers are considered from two main perspectives including the type of application domain and the utilized computing paradigm. The proposed framework is illustrated in Figure 1.

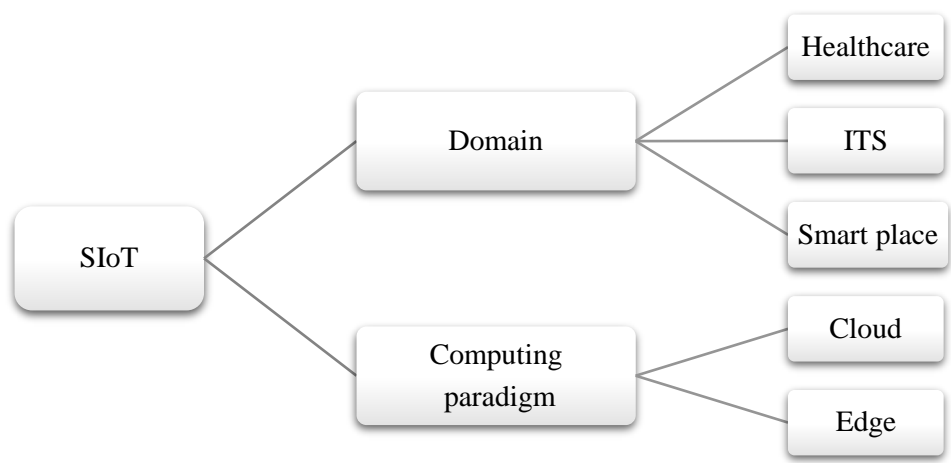

*Fig 1. The proposed framework*

Domain type depicts the main application of the considered SIoT system. IoT has previously entered almost all domains of daily lives. SIoT has limitations to be applied in all these domains, because of the social nature of it. In general, SIoT systems have been integrated into a few fields such as Healthcare (Baktha, Dev, Gupta, Agarwal, & Balamurugan, 2017; Poorejbari & Vahdat-Nejad, 2014) and Intelligent Transportation Systems ( Alam, Saini, & El Saddik, Toward social internet of vehicles: Concept, architecture, and applications, 2015; Vahdat-Nejad, Ramazani, Mohammadi, & Mansoor, 2016) to facilitate the process of automation and intelligence.

In recent decades, different computing paradigms have been introduced such as cluster (Barak & La'adan , 1998), grid (Yu & Buyya, 2005), pervasive (Satyanarayanan, 2001), cloud (Armbrust, et al., 2010), edge (Shi, Cao, Zhang, Li, & Lanyu Xu, 2016) and fog (Yi, Li, & Li, 2015) computing. Owing to the fact that the Internet of Things is arisen from these computing paradigms, SIoT solutions sometimes have employed these computing paradigms to fulfil the SIoT idea. In the proposed framework, the existing articles are also investigated based on the type of their computing paradigm. The advantage of such an investigation is to forecast the future integrated computing paradigm.

Table 1 demonstrates the overall specifications of the investigated papers. Since this survey is carried out in the beginning of 2018 and due to the required time for reviewing and publishing a paper, most of the considered papers are published in 2017. It shows the novelty of the SIoT paradigm. Although numerous papers have been published either in IoT or social computing, the number of outstanding works done on SIoT is limited. In the following sections, these projects are investigated and classified from the perspectives of domain and computing paradigm.

*Table 1. Projects overview*

| Project | Reference | Country | Year | Publication | | Publisher |
|---|---|---|---|---|---|---|
| | | | | Conference | Journal | |
| EMS | (Miori & Russo, 2017) | Italy | 2017 | * | | IEEE |
| V-doctor | (Ruggeri & Briante, 2017) | Italy | 2017 | * | | IEEE |
| PHY | (Hao & Wang, 2017) | Canada | 2017 | * | | IEEE |
| VSNP | (Jain, Brar, Malhotra, Rani, & Ahmed, 2018) | India | 2018 | | * | Elsevier |
| tNote | (Alam, Saini, & El Saddik, tNote: A Social Network of Vehicles under Internet of Things, 2014) | Canada | 2014 | * | | Springer |
| SIoV | (Nitti, Girau, Floris, & Atzori, 2014) | Italy | 2014 | * | | IEEE |
| LSE | (Cicirelli, et al., An edge-based approach to develop large-scale smart environments by leveraging SIoT, 2017) | Italy | 2017 | * | | IEEE |
| HVAC | (Marche, Nitti, & Pilloni, 2017) | Italy | 2017 | * | | IEEE |
| iSapiens | (Cicirelli, et al., iSapiens: A platform for social and pervasive smart environments, 2016) | Italy | 2016 | * | | IEEE |
| COSMOS | (Voutyras, Bourelos, Kyriazis, & Varvarigou, 2014) | Greece | 2014 | * | | IEEE |
| I-Painting | (Fu, et al., 2017) | China | 2017 | * | | Springer |

## 3. PROJECT REVIEW: DOMAIN

Internet of Things has entered into various fields of human life (Gubbi, Buyya, Marusic, & Palaniswami, 2013). Comparing to IoT, SIoT uses the social structure of things and has limitations in the application domain. After reviewing previous researches on SIoT from the perspective of application domain, they can be classified into the groups of Healthcare, Intelligent Transportation System and Smart place. In the following, these domains are discussed:

Healthcare: Healthcare domain has long been one of the most momentous concerns of humanity. Today, SIoT has gained much significance in this domain ( Riazul Islam, Kwak, & Humaun Kabir, 2015). It is necessary to provide a system for elderly care as senility abruptly affects the health and many of the elderly suffer from at least one chronic illness. Among the research studies, EMS project introduces an SIoT-based elderly monitoring system, which collects and analyzes environmental and physical data and sends them to the caregivers as needed. In the v-doctor project, a framework for the integration of E-health and SIoT has been proposed to develop the monitoring system of elderly. This system can provide special services for elderly through the issuance of medical guidelines and by discovering nearby objects and choosing the people who can help them. The PHY-Aided project introduces a security technique for protecting the healthcare system of SIoT. In this system, social networks are employed as a trustworthy platform for sharing the data of users and healthcare providers.

Intelligent Transportation System(ITS): Transportation is another domain that IoT plays an important role in ( Antonio Guerrero-ibanez, Sherali Zeadally, & Contreras-Castillo, 2015). Nowadays, the use of vehicles has increased dramatically as regards to the ongoing expansion of cities and population growth. It leads to the increase of traffic, accident and air pollution. Accordingly, transportation is one of the most important domains that benefits from the advantages of IoT to solve the mentioned problems (Manihatty Bojan, Raman Kumar, & Manihatty Bojan, 2014). In this regard, IoV has been proposed earlier (Whaiduzzaman, Sookhak, Gani, & Buyya, 2014; Yang, Wang, & Li, 2014) and its various social structures and types of communication have been investigated (Kaiwartya, et al., 2016). The importance of transportation domain is to the extent that it has also been penetrated by the paradigm of SIoT and several studies have been conducted in this field (P George, et al., 2017). Indeed, SIoT is introduced to omit the main difficulties of traffic management (Alam, Saini, & Saddik, Toward Social Internet of Vehicles: Concept, Architecture, and Applications, 2015). Among the investigated projects, SIoV consists of three components including management for collecting and analyzing data, security for managing the trust, and facilities to discover desired services for Intelligent Transportation Systems.

Vehicular Social Networking(VSN) is a type of social networks built by users on the roads. SIoV is applied in the VSNP project for the interaction of vehicles and raising the level of driving knowledge. Afterwards, an algorithm is proposed to control the traffic and road safety. In tNote, a Social Networking Vehicle Architecture (SIoV) and an infrastructure for data storage on VANET are proposed, which allow users to share their information with other vehicles.

Smart Place: Smart services and applications can affect and improve the quality of our daily lives. Smart Place is a concept that allows the use of advanced technologies for urban environments and buildings. It results in welfare in the quality of life by providing a variety of services (Rashidi, J. Cook, B. Holder, & Schmitter-Edgecombe, 2011). Smart building is a solution for monitoring and managing a building. It has various capabilities such as temperature, lighting and gate control as well as safety (Risteska Stojkoska & Trivodaliev, 2017).

Among the observed projects, the LSE introduces an SIoT-based approach to address the security issues such as trust, entity discovery, and interoperability in large-scale smart environments. In this system, an intermediate layer is utilized in order to hide the heterogeneity and mobility of devices in the smart place. The HVAC project proposes a smart system for efficient consumption of energy in smart buildings and employs the SIoT paradigm to reduce the cost of consuming electricity and energy. Similarly, to diminish the energy consumption, the COSMOS project provides a platform for handling limited resources in the SIoT paradigm. In this system, users are able to receive an energy plan based on their personal needs and budget, which is obtained by observing the past energy consumption patterns on the COSMOS platform. Finally, a Java-based platform for designing and implementing the smart environment is introduced by the iSapiens, which uses SIoT to address the constraints of scalability and computing capacity. Table 2 categorizes the considered projects regarding the domain type.

*Table 2. Project review according to domain type*

| Project | Reference | Domain | | |
|---|---|---|---|---|
| | | Healthcare | ITS | Smart Place |
| EMS | (Miori & Russo, 2017) | * | | |
| V-doctor | (Ruggeri & Briante, 2017) | * | | |
| PHY | (Hao & Wang, 2017) | * | | |
| VSNP | (Jain, Brar, Malhotra, Rani, & Ahmed, 2018) | | * | |
| t-Note | (Alam, Saini, & El Saddik, tNote: A Social Network of Vehicles under Internet of Things, 2014) | | * | |
| SIoV | (Nitti, Girau, Floris, & Atzori, 2014) | | * | |
| LSE | (Cicirelli, et al., An edge-based approach to develop large-scale smart environments by leveraging SIoT, 2017) | | | * |
| HVAC | (Marche, Nitti, & Pilloni, 2017) | | | * |
| iSapiens | (Cicirelli, et al., iSapiens: A platform for social and pervasive smart environments, 2016) | | | * |
| COSMOS | (Voutyras, Bourelos, Kyriazis, & Varvarigou, 2014) | | | * |

## 4. PROJECT REVIEW: COMPUTING PARADIGM

Several computing paradigms have been proposed in recent decades and evolved over time (Vahdat-Nejad, Izadpanah, & Ostadi-Eilaki, Context-aware cloud-based systems: design aspects, 2017; Vahdat-Nejad, Ostadi Eilaki, & Izadpanah, Towards a Better Understanding of Ubiquitous Cloud Computing, 2018). In this section, the latest computing paradigms that are used in SIoT are investigated. In general, considered research projects are divided in two categories based on the employed computing paradigm: cloud and edge computing. In the following, these projects are investigated.

Cloud Computing: The goal of cloud computing is to provide diverse computing services via the Internet. In fact, computing services are supplied to the users as web services (Prasad Rimal, Choi, & Lumb, 2009; Zhang, Cheng, & Boutaba, 2010; Rahimi, Ren, Liu, V. Vasilakos, & Venkatasubramanian, 2014). According to the NIST definition "cloud computing is a model for enabling ubiquitous, convenient, and on-demand network access to a shared pool of configurable computing resources (e.g., networks, servers, storage, applications and services) that can be rapidly provisioned and released with minimal management effort or service provider interaction" (Mel & Grance, 2011). Cloud computing users are not commonly the owners of the cloud infrastructure; however they lease it out from a third-party supplier to avoid high costs ( Chaisiri, Lee, & Niyato, 2012). Indeed, resources are used by consumers in the form of a service and users only pay the renting cost of the resources they have utilized (Al-Roomi, Al-Ebrahim, Buqrais , & Ahmad, 2013). Some of other advantages of cloud computing include cost effectiveness and enhanced efficiency, reliability, security and scalability (Wang, Wang, Ren, & Lou, Privacy-Preserving Public Auditing for Data Storage Security in Cloud Computing, 2010; Zissis & Lekkas, 2012; Fernando, W.Loke, & Rahayu, 2013; Mollah, Azad, & Vasilakos, 2017).

Generally, the investigated projects employ cloud due to the storage, processing and concurrent service provisioning capabilities. Owing to the limited storage of mobile devices of users, cloud is employed as a storage resource in most of the projects (Wang, Wang, Ren, & Lou, Privacy-Preserving Public Auditing for Data Storage Security in Cloud Computing, 2010; Wang, Wang, & Ren, Toward Secure and Dependable Storage Services in Cloud Computing, 2012).

Regarding the investigated projects, a cloud-based SIoT system is provided in the V-doctor system. Cloud is used in this project to improve scalability since computational load is increased by unlimited requests for simultaneous searches from different devices. In the VSNP project, vehicle data is uploaded, stored and processed on the cloud to control the traffic. Likewise, to manage and process the data of vehicles and road side units, cloud is applied in tNote project. I-painting system, which provides a smart painting service for children utilizes cloud for storing and sharing the paintings. Furthermore, a cloud-based system to process and store data in smart buildings is introduced in HVAC project. To diminish the energy consumption, the COSMOS project provides a method commensurate with limited resources of SIoT, which uses cloud to store data.

Edge Computing: In cloud computing, the complexity of processing, storing and network configuration is hidden in data centers. Nevertheless, cloud is not proper for real-time applications due to the fact that it is far away from the user's devices and has a WAN delay (Yu, Liang, & He, 2017; Mao, You, & Zhang, 2017). On the other hand, regarding the ever-increasing use of IoT as well as mobile devices and the huge volume of data that sensors and devices produce, real-time processing of this information has become a challenge (Abbas, Zhang, & Taherkordi, 2018). Consequently, a novel paradigm is recently introduced, which is called edge computing. It accomplishes storing and processing on the edge of the network (Ahmed & Rehmani, 2017).

Wide smart environments are inherently open and dynamic and include a large number of interactions. In these environments some factors such as trust, processing and managing data, interoperability and scalability are momentous. To manage and process data, LSE project employs cloud computing paradigm integrated with edge computing. The edge computing paradigm along with the cloud provides a great computational power for processing tasks. Likewise, the iSapiens system improves the storing and distributed computing by the help of edge computing and also eliminates the side-effects such as transmission delay and bandwidth shortage. Table 3 summarizes these research papers in terms of the type of utilized computing paradigm.

*Table 3. Project review according to computing paradigm*

| Project | Reference | Cloud Computing | | | Edge Computing | | |
|---|---|---|---|---|---|---|---|
| | | Storage | Processing | Serving simultaneously | Reducing Energy | Reducing Bandwidth | Reducing Latency |
| V-doctor | (Ruggeri & Briante, 2017) | | | * | | | |
| VSNP | (Jain, Brar, Malhotra, Rani, & Ahmed, 2018) | * | * | | | | |
| t-Note | (Alam, Saini, & El Saddik, tNote: A Social Network of Vehicles under Internet of Things, 2014) | | * | | | | |
| HVAC | (Marche, Nitti, & Pilloni, 2017) | * | * | | | | |
| I-Painting | (Fu, et al., 2017) | * | | | | | |
| COSMOS | (Voutyras, Bourelos, Kyriazis, & Varvarigou, 2014) | * | | | | | |
| LSE | (Cicirelli, et al., An edge-based approach to develop large-scale smart environments by leveraging SIoT, 2017) | | * | | | * | |
| iSapiens | (Cicirelli, et al., iSapiens: A platform for social and pervasive smart environments, 2016) | | | | * | * | * |

## 5. CONCLUSION

In this survey, the SIoT research studies have been investigated from two main perspectives, including the type of application domain and the utilized computing paradigm. Despite the fact that a few research studies have been conducted in the SIoT, their growth rate is strikingly high that makes it a hot research topic. Although SIoT has entered into various specialized domains such as Healthcare and Intelligent Transportation Systems, the number of projects in each of these fields is significantly low.

Regarding the investigated projects, it has been specified that in the majority of SIoT systems, cloud is employed for storage as well as simultaneous service provisioning due to limited resources of users' mobile devices. Besides, despite the novelty of edge computing, some research studies have exploited it to solve the SIoT realization challenges, which demonstrates the necessity of introducing a new architecture for SIoT based on new computing models. Indeed, SIoT is a new paradigm, still in its infancy, that tries to overcome the issues of scalability, reliability, and discovery of resources and information through inspiring from human social networks and provides a platform for better interactions of humans and things. It is expected to see a large number and more specialized research in this emerging field in the future.

**REFERENCES**

Aggarwal, C., Ashish, N., & Sheth, A. (2013). the internet of things: a survey from the data-centric perspective. In *Managing and Mining Sensor Data* (pp. 383-428). Springer.

Alam, K., Saini, M., & El Saddik, A. (2015). Toward social internet of vehicles: Concept, architecture, and applications. *Access, 3*, 343 - 357.

Antonio Guerrero-ibanez, J., Sherali Zeadally, S., & Contreras-Castillo, J. (2015). Integration challenges of intelligent transportation systems with connected vehicle, cloud computing, and internet of things technologies. *IEEE Wireless Communications , 22*(6), 122 - 128.

Chaisiri, S., Lee, B.-S., & Niyato, D. (2012). Optimization of Resource Provisioning Cost in Cloud Computing. *IEEE Transactions on Services Computing, 5*(2), 164 - 177.

Lin, Z., & Dong , L. (2018). Clarifying Trust in Social Internet of Things. *Transactions on Knowledge and Data Engineering, 30*(2), 234 - 248.

Rahimi, M., Ren, J., Liu, C. H., V. Vasilakos, A., & Venkatasubramanian, N. (2014). Mobile Cloud Computing: A Survey, State of Art and Future Directions. *Mobile Networks and Applications, 19*(2), 133–143.

Riazul Islam, S., Kwak, D., & Humaun Kabir, M. (2015). The Internet of Things for Health Care: A Comprehensive Survey. *IEEE Access , 3*, 678 - 708.

Truong, N., Lee, H., Askwith, B., & Lee, G. (2017). Toward a Trust Evaluation Mechanism in the Social Internet of Things. *Sensors, 17*(6).

Abbas, N., Zhang, Y., & Taherkordi, A. (2018). Mobile Edge Computing: A Survey. *IEEE Internet of Things Journal, 5*(1), 450 - 465.

Afzal, B., Umair, M., Shah, G., & Ahmed, E. (2018). Enabling IoT platforms for social IoT applications: Vision, feature mapping, and challenges. *Future Generation Computer Systems*.

Ahmed, E., & Rehmani, M. (2017). Mobile Edge Computing: Opportunities, solutions, and challenges. *Future Generation Computer Systems, 70*, 59-63.

Alam, K., Saini, M., & El Saddik, A. (2014). tNote: A Social Network of Vehicles under Internet of Things. *Internet of Vehicles – Technologies and Services* (pp. 227-236). Beijing, China: springer.

Alam, K., Saini, M., & Saddik, A. (2015). Toward Social Internet of Vehicles: Concept, Architecture, and Applications. *IEEE Access, 3*, 343 - 357.

Al-Fuqaha, A., Guizani, M., Mohammadi, M., Aledhari, M., & Ayyash, M. (2015). Internet of Things: A Survey on Enabling Technologies, Protocols, and Applications. *Communications Surveys & Tutorials, 17*(4), 2347 - 2376.

Al-Roomi, M., Al-Ebrahim, S., Buqrais , S., & Ahmad, I. (2013). Cloud Computing Pricing Models: A Survey. *International Journal of Grid and Distributed Computing, 6*(5), 93-106.

Armbrust, M., Fox, A., Griffith, R., Joseph, A., Katz, R., Konwinski, A., . . . Zaharia, M. (2010). A view of cloud computing. *Communications of the ACM, 53*(4), 50-58.

Atzori, L., Iera, A., & Morabito, G. (2011). SIoT: Giving a Social Structure to the Internet of Things. *Communications Letters, 15*(11), 1193 - 1195.

Atzori, L., Iera, A., Morabito, G., & Nitti, M. (2012). The Social Internet of Things (SIoT) – When social networks meet the Internet of Things: Concept, architecture and network characterization. *Computer Networks, 56*(16), 3594-3608.

Atzori, L., Iera, A., & Morabito, G. (2010). The Internet of Things: A survey. *Computer Networks, 54*(15), 2787-2805.

Baktha, K., Dev, M., Gupta, H., Agarwal, A., & Balamurugan, B. (2017). Social Network Analysis in Healthcare. In *Internet of Things and Big Data Technologies for Next Generation Healthcare* (pp. 309-344). springer.

Barak, A., & La'adan, O. (1998). The MOSIX multicomputer operating system for high performance cluster computing. *Future Generation Computer Systems, 13*(6), 361-372.

Cicirelli, F., Guerrieri, A., Spezzano, G., Vinci, A., Briante, O., & Ruggeri, G. (2016). iSapiens: A platform for social and pervasive smart environments. *3rd World Forum on Internet of Things (WF-IoT).* Reston, VA, USA.

Cicirelli, F., Guerrieri, A., Spezzano, G., Vinci, A., Briante, O., Iera, A., & Ruggeri, G. (2017). An edge-based approach to develop large-scale smart environments by leveraging SIoT. *14th International Conference on Networking, Sensing and Control (ICNSC).* Calabria, Italy.

Fernando, N., W.Loke, S., & Rahayu, W. (2013). Mobile cloud computing: A survey. *Future Generation Computer Systems, 29*(1), 84-106.

Fu, Z., Lin, J., Li, Z., Du, W., Zhang, J., & Ye, S. (2017). Intelligent painting based on social internet of things. *5th International Conference on Distributed, Ambient, and Pervasive Interactions* (pp. 335-346). Canada: Springer.

Ghane'i-Ostad, M., Vahdat-Nejad, H., & Abdolrazzagh-Nezhad, M. (2018). Detecting overlapping communities in LBSNs by fuzzy subtractive clustering. *Social Network Analysis and Mining*.

Girau, R., Martis, S., & Atzori, L. (2015). A Cloud-Based Platform of the Social Internet of Things. In *Internet of Things. IoT Infrastructures* (pp. 77-88). springer.

Gubbi, J., Buyya, R., Marusic, S., & Palaniswami, M. (2013). Internet of Things (IoT): A Vision, Architectural Elements, and Future Directions. *Future Generation Computer Systems, 29*(7), 1645-1660.

Hao, P., & Wang, X. (2017). A PHY-Aided Secure IoT Healthcare System With Collaboration of Social Networks. *Vehicular Technology.* Toronto, ON, Canada : IEEE.

Jain, B., Brar, G., Malhotra, J., Rani, S., & Ahmed, S. (2018). A cross layer protocol for traffic management in Social Internet of Vehicles. *Future Generation Computer Systems, 82*, 707-714.

Jara, A., Bocchi, Y., & Genoud, D. (2014). Social Internet of Things: The potential of the Internet of Things for defining human behaviours. *Intelligent Networking and Collaborative Systems.* Salerno, Italy.

Kaiwartya, O., Abdullah, A. h., Cao, Y., Altameem, A., Prasad, M., Lin, C.-T., & Liu, X. (2016). Internet of Vehicles: Motivation, Layered Architecture, Network Model, Challenges, and Future Aspects. *IEEE Access, 4*, 5356 - 5373.

Lee, H., & Kwon, J. (2015). Survey and Analysis of Information Sharing in Social IoT. *Disaster Recovery and Business Continuity.* Jeju, South Korea: IEEE.

Manihatty Bojan, T., Raman Kumar, U., & Manihatty Bojan, V. (2014). An internet of things based intelligent transportation system. *IEEE International Conference on Vehicular Electronics and Safety.* Hyderabad, India.

Mao, Y., You, C., & Zhang, J. (2017). A Survey on Mobile Edge Computing: The Communication Perspective. *IEEE Communications Surveys & Tutorials, 19*(4), 2322 - 2358.

Marche, C., Nitti, M., & Pilloni, V. (2017). Energy efficiency in smart building: a comfort aware approach based on Social Internet of Things. *Global Internet of Things Summit (GIoTS).* Geneva, Switzerland.

Mel, P., & Grance, T. (2011). *The NIST Definition of Cloud.* United States: National Institute of Standards and Technology.

Miorandi, D., Sicari, S., De Pellegrini, F., & Chlamtac, I. (2013). Internet of things: Vision, applications and research challenges. *Ad Hoc Networks, 29*(7), 1645-1660.

Miori, V., & Russo, D. (2017). Improving life quality for the elderly through the Social Internet of Things (SIoT). *Global Internet of Things Summit.* Geneva, Switzerland .

Mollah, M., Azad, M. A., & Vasilakos, A. (2017). Security and privacy challenges in mobile cloud computing: Survey and way ahead. *Journal of Network and Computer Applications, 84*, 38-54.

Nitti, M., Atzori, L., & Cvijikj, I. (2014). Network navigability in the social Internet of Things. *World Forum on Internet of Things.* Seoul, South Korea.

Nitti, M., Girau, R., & Atzori, L. (2014). Trustworthiness Management in the Social Internet of Things. *IEEE Transactions on Knowledge and Data Engineering, 26*(5), 1253 - 1266.

Nitti, M., Girau, R., Atzori, L., Iera, A., & Morabito, G. (2012). A subjective model for trustworthiness evaluation in the social Internet of Things. *Personal Indoor and Mobile Radio Communications.* Sydney, NSW, Australia: IEEE.

Nitti, M., Girau, R., Floris, A., & Atzori, L. (2014). On adding the social dimension to the Internet of Vehicles: friendship and middleware. *International Black Sea Conference on Communications and Networking (BlackSeaCom).* Odessa, Ukraine: IEEE.

P George, S., Wilson, N., U Nair, K., Michael, K., B Aricatt, M., & George K, T. (2017). Social Internet of Vehicles. *International Research Journal of Engineering and Technology (IRJET), 4*(4), 712-717.

Preprint of the book-chapter submitted for publication in the book entitled "Toward Social Internet of Things (SIoT): Enabling Technologies, Architectures and Applications", Springer, 2019
You can find the published version here: https://link.springer.com/chapter/10.1007/978-3-030-24513-9_8
Poorejbari, S., & Vahdat-Nejad, H. (2014). An Introduction to Cloud-Based Pervasive Healthcare Systems. *3rd International Conference on Context-Aware Systems and Applications.* Dubai, United Arab Emirates: ACM.

Prasad Rimal, B., Choi, E., & Lumb, I. (2009). A Taxonomy and Survey of Cloud Computing Systems. *Fifth International Joint Conference on INC, IMS and IDC.* Seoul, South Korea.

Rabadiya, K., Makwana, A., & Jardosh, S. (2017). Revolution in Networks of Smart Objects: Social Internet of Things. *Soft Computing and its Engineering Applications.* Changa, India.

Rashidi, P., J. Cook, D., B. Holder, L., & Schmitter-Edgecombe, M. (2011). Discovering Activities to Recognize and Track in a Smart Environment. *IEEE Transactions on Knowledge, 23*(4), 527 - 539.

Risteska Stojkoska, B., & Trivodaliev, K. (2017). A review of Internet of Things for smart home: Challenges and solutions. *Journal of Cleaner Production, 140*(3), 1454-1464.

Ruggeri, G., & Briante, O. (2017). A framework for IoT and E-Health Systems Integration based on the Social Internet of Things Paradigm. *International Symposium on Wireless Communication Systems (ISWCS).* Bologna, Italy.

Satyanarayanan, M. (2001). Pervasive computing: Vision and challenges. *Personal Communications, 8*(4), 10 - 17.

Shi, W., Cao, J., Zhang, Q., Li, Y., & Lanyu Xu. (2016). Edge Computing: Vision and Challenges. *Internet of Things Journal, 3*(5), 637 - 646.

Stankovic, J. (2014). Research Directions for the Internet of Things. *IEEE Internet of Things Journal, 1*(1), 3-9.

Vahdat-Nejad, H., Izadpanah, S., & Ostadi-Eilaki, S. (2017). Context-aware cloud-based systems: design aspects. *Cluster Computing*, 1-17.

Vahdat-Nejad, H., Ostadi Eilaki, S., & Izadpanah, S. (2018). Towards a Better Understanding of Ubiquitous Cloud Computing. *Cloud Applications and Computing, 8*(1).

Vahdat-Nejad, H., Ramazani, A., Mohammadi, T., & Mansoor, W. (2016). A survey on context-aware vehicular network applications. *Vehicular Communications, 3*, 43-57.

Voutyras, O., Bourelos, P., Kyriazis, D., & Varvarigou, T. (2014). An architecture supporting knowledge flow in Social Internet of Things systems. *Wireless and Mobile Computing, Networking and Communications.* Larnaca, Cyprus.

Wang, C., Wang, Q., Ren, K., & Lou, W. (2010). Privacy-Preserving Public Auditing for Data Storage Security in Cloud Computing. *Proceedings IEEE INFOCOM.* San Diego, CA, USA.

Wang, C., Wang, Q., & Ren, K. (2012). Toward Secure and Dependable Storage Services in Cloud Computing. *IEEE Transactions on Services Computing, 5*(2), 220 - 232.

Whaiduzzaman, M., Sookhak, M., Gani, A., & Buyya, R. (2014). A survey on vehicular cloud computing. *Journal of Network and Computer Applications, 40*, 325-344.

Whitmore, A., Agarwal, A., & Da Xu, L. (2015). The Internet of Things—A survey of topics and trends. *Information Systems Frontiers, 17*(2), 261–274.

Xiao, H., Sidhu, N., & Christianson, B. (2015). guarantor and reputation based trust model for social internet of things. *Wireless Communications and Mobile Computing.* Dubrovnik, Croatia.

Yang, F., Wang, S., & Li, J. (2014). An Overview of Internet of Vehicles. *IEEE, 11*(10), 1 - 15.

Yi, S., Li, C., & Li, Q. (2015). Survey of Fog Computing: Concepts, Applications and Issues. *Mobidata '15 Workshop on Mobile Big Data* (pp. 37-42). Hangzhou, China: ACM.

Yu, J., & Buyya, R. (2005). A Taxonomy of Workflow Management Systems for Grid Computing. *Grid Computing, 3*(3-4), 171-200.

Yu, W., Liang, F., & He, X. (2017). A Survey on the Edge Computing forthe Internet of Things. *IEEE Access, 6*, 6900 - 6919.

Zargari Asl, H., Iera, A., Atzori, L., & Morabito, G. (2013). How often social objects meet each other? Analysis of the properties of a social network of IoT devices based on real data. *Global Communications Conference.* Atlanta, GA, USA.

Zhang, Q., Cheng, L., & Boutaba, R. (2010). Cloud computing: state-of-the-art and research challenges. *Journal of Internet Services and Applications, 1*(1), 7–18.

Zissis, D., & Lekkas, D. (2012). Addressing cloud computing security issues. *Future Generation Computer Systems, 28*(3), 583-592.